\documentclass[sigconf]{acmart}

\AtBeginDocument{%
  \providecommand\BibTeX{{%
    \normalfont B\kern-0.5em{\scshape i\kern-0.25em b}\kern-0.8em\TeX}}}

\usepackage{array}
\usepackage{graphicx}
\usepackage[switch]{lineno}
\usepackage{amsmath}

\usepackage{amssymb}
\usepackage{multirow}
\usepackage{booktabs}

\usepackage{algorithm}
\usepackage{algorithmic}
\usepackage{xcolor}
\usepackage{xspace}

\newcommand{\paratitle}[1]{\vspace{1.5ex}\noindent\textbf{#1}}
\newcommand{\ie}{\emph{i.e.,}\xspace}

\newcommand{\eg}{\emph{e.g.,}\xspace}
\newcommand{\baby}{\textsc{CoHHN}\xspace}
\newcommand{\babyuni}{\textsc{CoHHN}}

\copyrightyear{2022}
\acmYear{2022}
\setcopyright{acmcopyright}
\acmConference[SIGIR '22]{Proceedings of the 45th International ACM SIGIR Conference on Research and Development in Information Retrieval}{July 11--15, 2022}{Madrid, Spain.}
\acmBooktitle{Proceedings of the 45th International ACM SIGIR Conference on Research and Development in Information Retrieval (SIGIR '22), July 11--15, 2022, Madrid, Spain}
\acmPrice{15.00}
\acmISBN{978-1-4503-8732-3/22/07}
\acmDOI{10.1145/3477495.3532043}

\begin{document}
\fancyhead{}

\title{Price DOES Matter! Modeling Price and Interest Preferences in Session-based Recommendation}



\author{Xiaokun Zhang}
\affiliation{%
  \institution{Dalian University of Technology}
}
\email{kun@mail.dlut.edu.cn}

\author{Bo Xu}
\affiliation{%
  \institution{Dalian University of Technology}
  }
\email{xubo@dlut.edu.cn}

\author{Liang Yang}
\affiliation{%
  \institution{Dalian University of Technology}
  }
\email{liang@dlut.edu.cn}

\author{Chenliang Li}
\affiliation{%
  \institution{Wuhan University}
  }
\email{cllee@whu.edu.cn}

\author{Fenglong Ma}
\affiliation{%
  \institution{Pennsylvania State University}
  }
\email{fenglong@psu.edu}

\author{Haifeng Liu}
\affiliation{%
  \institution{Dalian University of Technology}
  }
\email{liuhaifeng@mail.dlut.edu.cn}

\author{Hongfei Lin}
\authornote{Corresponding Author.}
\affiliation{%
  \institution{Dalian University of Technology}
  }
\email{hflin@dlut.edu.cn}

\renewcommand{\shortauthors}{Xiaokun Zhang, et al.}
\begin{abstract}

Session-based recommendation aims to predict items that an anonymous user would like to purchase based on her short behavior sequence. The current approaches towards session-based recommendation only focus on modeling users' interest preferences, while they all ignore a key attribute of an item, \ie the price. Many marketing studies have shown that the price factor significantly influences users' behaviors and the purchase decisions of users are determined by both price and interest preferences simultaneously. However, it is nontrivial to incorporate price preferences for session-based recommendation. Firstly, it is hard to handle heterogeneous information from various features of items to capture users' price preferences. Secondly, it is difficult to model the complex relations between price and interest preferences in determining user choices.

To address the above challenges, we propose a novel method Co-guided Heterogeneous Hypergraph Network (\baby) for session-based recommendation. Towards the first challenge, we devise a heterogeneous hypergraph to represent heterogeneous information and rich relations among them. A dual-channel aggregating mechanism is then designed to aggregate various information in the heterogeneous hypergraph. After that, we extract users' price preferences and interest preferences via attention layers. As to the second challenge, a co-guided learning scheme is designed to model the relations between price and interest preferences and enhance the learning of each other. 
Finally, we predict user actions based on item features and users' price and interest preferences. 
Extensive experiments on three real-world datasets demonstrate the effectiveness of the proposed \baby. Further analysis reveals the significance of price for session-based recommendation.

\end{abstract}



\maketitle

\section{Introduction}

Recommender system (RS), as an effective tool to mitigate information explosion, plays a critical role in modern e-commerce systems. 
Conventional RS usually uses long-term behaviors of a user to predict her future actions accordingly~\cite{IKNN,Koren@Computer2009}. In many cases, however, the user's portrait and her rich historical behavior data are unavailable due to privacy policy or non-logged-in nature. To address this issue, session-based recommendation (SBR) is proposed to predict actions of an anonymous user based on limited interactions (\eg clicked items and their associated features) within the current session \cite{NARM, SR-GNN, ludewig@UMUAI2020}.

Existing SBR approaches focus on modeling the user's \textbf{interest preferences}, \ie how much a user likes an item, usually based on recurrent neural network (RNN)~\cite{GRU4Rec, NARM}, attention network~\cite{SASRec,STAMP} and graph neural network (GNN)~\cite{SR-GNN,LESSR}. Although having achieved impressing performance, these approaches all ignore a significantly important factor, \ie the user's \textbf{price preferences}, which aim to describe how much money a user is willing to pay for an item. Many marketing studies have shown that users' buying behaviors are strongly influenced by the price factor~\cite{chen@1998,umberto2015price,han@2001}. Thus, the price preferences of users should be taken into consideration when predicting their actions. However, we face two main challenges when modeling users' price preferences for SBR.

In fact, price preferences of a user are not fixed, which dramatically vary according to item categories~\cite{umberto2015price,Zheng@ICDE2020}. 
For instance, a user may purchase an \emph{expensive laptop} out of high requirements for the performance of computing, while she might buy \emph{affordable pajamas} to wear at home. 
Thus, when incorporating users' price preferences, we need to take the corresponding categories of items into account. 
In such a way, multiple types of information will be involved to describe users' historical behaviors, including \emph{a series of items}, \emph{item prices}, and \emph{item categories}. Those \textbf{heterogeneous information} may make existing approaches such as RNN \cite{GRU4Rec, NARM}, attention network \cite{SASRec,STAMP}, and GNN \cite{SR-GNN,LESSR} failed since they are designed to only capture a single type of information. 
Therefore, to model price preferences for SBR, the first challenge is how to handle such heterogeneous information.

Besides, there is a phenomenon called \emph{price elasticity} in economics, which means that the money that a user is willing to pay for an item fluctuates according to her interest in it \cite{brynjolfsson2000price}. This phenomenon clearly shows that the buying behaviors of a user are determined by both price and interest preferences, which is intuitive since users often make price-interest trade-offs when they purchase items. 
For example, a user is likely to purchase an expensive item out of strong interest, even if the price of the item exceeds her expectations. In this case, the user adjusts her price preferences due to strong interest. Similarly, users will also buy some items that are not their favorite because of its low price. Therefore, in SBR, \textbf{both price and interest preferences are indispensable}. However, how to model the complex relations between price and interest preferences is another challenge.

To tackle the aforementioned challenges, we propose a novel model called \underline{Co}-guided \underline{H}eterogeneous \underline{H}ypergraph \underline{N}etwork (\baby) for SBR. 
In particular, for addressing the first challenge, an intuitive way is to directly apply state-of-the-art heterogeneous graph-based approaches to model \emph{heterogeneous information}.
However, existing heterogeneous graph-based methods~\cite{Sun@KDD2012,Hu@EMNLP2019} can only capture up to $k$-hop pairwise relations (\ie \emph{item} $\to$ \emph{category}), where $k$ is always set to a small value (\eg 3) due to the over-smoothing issue \cite{LESSR}. 
In our setting, the relations among various information are high-order not pairwise such as the relations among $<$\emph{item}, \emph{price}, \emph{category}$>$. Such complex high-order dependencies may make general heterogeneous graph-based approaches failed. 
Recently, hypergraphs with degree-free hyperedges are proposed to encode high-order data correlation~\cite{Yadati@NIPS2019}. 
Inspired by both heterogeneous graphs and hypergraphs, in this paper, we propose a \textbf{heterogeneous hypergraph network}, which combines the advantages of heterogeneous graphs in modeling heterogeneous information and hypergraphs in capturing complex high-order dependencies, to handle heterogeneous information in SBR.

Specifically, we encode the following heterogeneous nodes in the heterogeneous hypergraph: item price, item ID and item category, all of which are closely related to price and interest preferences. We define three types of hyperedges, \ie feature hyperedge, price hyperedge and session hyperedge, which are used to represent multi-type relations among various nodes. Then a dual-channel aggregating mechanism is devised to propagate node embeddings via feature hyperedges. Based on the learned node embeddings, we apply attention layers to extract original price and interest preferences via price and session hyperedges, respectively.

As we mentioned in the second challenge, both price and interest preferences determine user behaviors. To address this challenge, we propose a novel \textbf{co-guided learning schema}, which can model the complex relations between price and interest preferences and enhance the learning of each other. Concretely, based on the original price and interest preferences learned from the heterogeneous hypergraph network, we further make these two preferences guide the learning of each other to enrich their semantics. Finally, \baby makes the recommendation based on item features and users' price and interest preferences.

In summary, the contributions of our work are as follows:

\vspace{-\topsep}
\begin{itemize}
    \item We highlight the significance of the price factor in determining user behaviors and take both price and interest preferences into consideration to predict user actions in SBR. To our best knowledge, we are the first to consider the price factor in the task of SBR.
    \item We propose a novel method called Co-guided Heterogeneous Hypergraph Network (\baby) to model price preferences, interest preferences and the relations between them to further boost the performance of SBR.
    \item Extensive experiments over three public benchmarks demonstrate the superiority of our proposed \baby compared with state-of-the-art approaches. Further analysis also justifies the importance of the price factor in SBR.
\end{itemize}

\section{Related Work}

\subsection{Session-based Recommendation} 
Session-based recommendation (SBR) has attracted increasing attention recently due to its highly practical value \cite{Wang@CS2021}. With the powerful representation ability, many neural models are applied to improve SBR. Most early works \cite{GRU4Rec,Tan@DLRS2016,NARM,P-RNN,HidasiK@CIKM2018,CSRM} use RNN and its variants by considering RNN's capability of processing sequential data. 
Attention mechanism is also commonly used in SBR to distinguish the importance of different items in a session \cite{ATEM,STAMP,SASRec,BERT4Rec,Chen@CIKM2019,DPAN}. 
Furthermore, there are some methods~\cite{FGNN,MTD,GCE-GNN,Star@CIKM2020} utilizing GNN to capture pairwise relations between items.
LESSR~\cite{LESSR} further handles information loss of GNN for SBR. 
In addition, some methods explore the characteristics of user behaviors in SBR, such as local invariance property~\cite{LINet}, multi interest~\cite{MCPRN,Tan@WSDM2021}, uncertainty~\cite{RAP, DIDN} and repeated consumption~\cite{RepeatNet,SLIST}.
Recently, SHARE \cite{Wang@SDM2021} and $S^2$-DHCN \cite{DHCN} introduce hypergraph in the task to model the high-order relations among items. Other methods augment session data by co-training \cite{Xia@CIKM2021} or counterfactual schema \cite{Wang@SIGIR2021}. 
Although price is an important factor in determining user behaviors, to our best knowledge, none of the existing works considers it in SBR.

\subsection{Price-aware Recommendation} 
Price is often a key consideration for purchases, but there is a few works introducing price factor into recommender system. \citet{schafer1999price} are pioneers to explore the impact of price factor on user behaviors. 
\citet{SIGIR2014price} use the price to hinder performance degradation caused by the unseen categories.
\citet{RecSys2020price} re-rank recommendation results by inferring the user affinity to price levels.
PUP \cite{Zheng@ICDE2020} builds a price-contained heterogeneous graph to model user purchase behaviors.  
These works only used the price factor as an auxiliary feature in modeling user actions, while ignoring the decisive role of price preferences in the user decision-making process. Besides, none of them is designed for session scenarios.

\subsection{Heterogeneous Graph and Hypergraph} 
Heterogeneous graph has been demonstrated effective in handling heterogeneous information~\cite{Sun@KDD2012,Hu@EMNLP2019}. 
For example, \citet{Fan@KDD2019} propose a metapath-guided method in heterogeneous graph for intent recommendation. 
Hypergraph is a graph in which a hyperedge can connect more than two nodes \cite{Zhou@NIPS2006,Yadati@NIPS2019}. This unique feature enables it to capture complex high-order dependencies among nodes, particularly in recommendation tasks. 
HyperRec \cite{Wang@SIGIR2020} utilizes this capacity of hypergraph to model the short-term user preferences for next-item recommendation.   
In this paper, we propose a novel heterogeneous hypergraph network which combines the advantages of the heterogeneous graph and hypergraph. The customized structure enables our proposed \baby to model the complex high-order dependencies among heterogeneous information and extract users' price and interest preferences more precisely.

\section{Preliminaries}
\subsection{Problem Statement}
The focus of our work is to model both price and interest preferences of users for improving session-based recommendation. Let $\mathcal{I}$ denote the set of all unique items and $|\mathcal{I}| = n$ is the total number of items. An item $x_i$ $\in$ $\mathcal{I}$ contains some features, including item ID, price and category. $\mathcal{S}$ = [$x_1, x_2, ..., x_m$] is a session produced by an anonymous user, where $m$ is the length of $\mathcal{S}$. The proposed \baby can be viewed as a function $f$ whose input is the current session $\mathcal{S}$. The output is \textbf{y} = $f$($\mathcal{S}$) = [$y_1, y_2, ..., y_n$], where $y_j$ ($j \in [1,..., n]$) indicates the likelihood that the user will interact with item $x_j$ next. 

\subsection{Price Discretization}

\begin{figure}[t]
  \centering
  \includegraphics[width=0.8\linewidth]{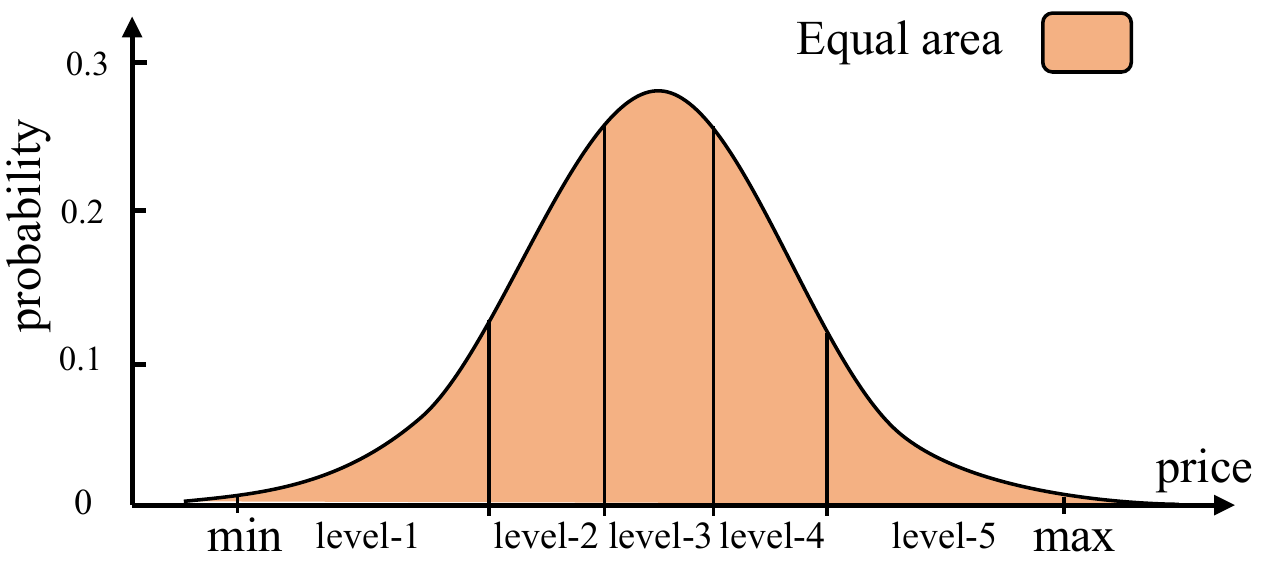}
  \caption{Logistic distribution. We discretize absolute price into different price levels by equally dividing price probability distribution in a category.}\label{logisticDistri}
\end{figure}

Absolute price of an item is not sufficient to determine if the item is expensive or not \cite{RecSys2020price}. For example, \$200 may be a bargain for a cellphone, but the same price would be very high for a lamp. Therefore, we have to discretize item price into different price levels according item categories, such that they can be compared across different categories. As indicated in \cite{KBS2018price}, the price distribution on a certain category is closer to the logistic distribution instead of widely used uniform distribution. As plotted in Figure~\ref{logisticDistri}, the probability density function of logistic distribution presents a high preference in the middle and low preference on both sides. Where price is concentrated, users have more items with similar price to choose, leading to higher price sensitivity and the price levels should be divided more finely, and vice versa. Furthermore, to balance the training data, the number of items contained in each price level should be in the same scale. Therefore, as shown in Figure~\ref{logisticDistri}, we discretize price into $\rho$ levels (\eg $\rho=5$), where the probability corresponding to each interval is equal. Formally, for an item $x_i$ with price $x_p$ and the price range of its category is $[min, max]$, we determine its price level as follows,
\begin{align}
    p_i &= \lfloor\frac{\Phi(x_p)-\Phi(min)}{\Phi(max)-\Phi(min)}\times\rho\rfloor
\end{align}

where $\Phi(x)$ is the cumulative distribution function of logistic distribution, which can be defined as follows,
\begin{align}
    \Phi(x) &= P(X \leq x) = \frac{1}{1 + e^{-\pi\frac{x-\mu}{\sqrt{3}\delta}} }
\end{align}
where $\mu$ and $\delta$ are expected value and standard deviation respectively. 

\subsection{Heterogeneous Hypergraph Construction}
\begin{figure*}[t]
  \centering
  \includegraphics[width=0.95\linewidth]{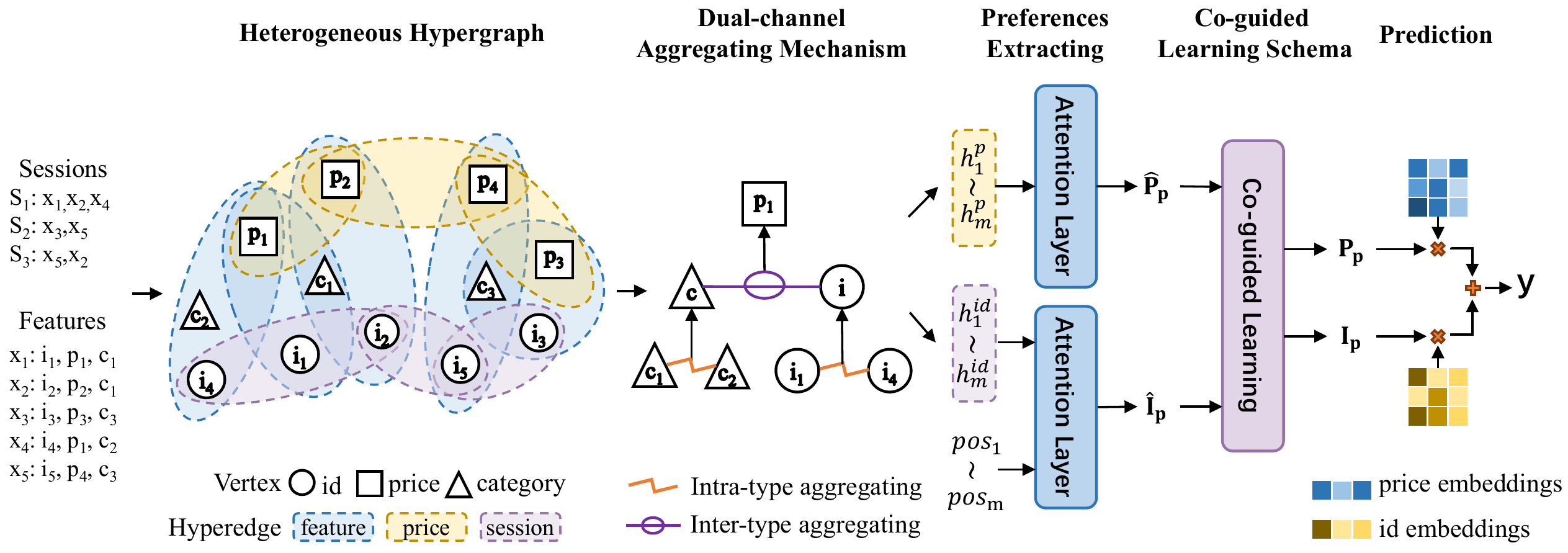}
  \caption{The workflow of the proposed \baby. We first construct a heterogeneous hypergraph based on all sessions. And a dual-channel aggregating mechanism is devised to aggregate various information among heterogeneous nodes. After that, we utilize attention layers to extract user price and interest preferences based on price and session hyperedges, respectively. A co-guided learning schema is then proposed to model the complex relations between these two preferences. Lastly, we predict user future actions based on item features and users' price and interest preferences. (\textit{Best viewed in color})}\label{CoHHN}
\end{figure*}

To extract users' price and interest preferences from heterogeneous information, we devise a customized heterogeneous hypergraph $\mathcal{G}$ = $(V, E)$, where $V$ is the set of all nodes and $E$ is the set of all hyperedges. Each node $ \mathbf{v}^{\tau} \in V$ owns a type $\tau$ and the nodes with same type form a set of homogeneous nodes $V^{\tau}$. The types of nodes we encode include item price ($V^{p}$), item ID ($V^{id}$) and item category ($V^{c}$), all of which are closely related with both price and interest preferences, such that $V = V^{p} \bigcup V^{id} \bigcup V^{c}$. Each hyperedge $\epsilon \in E$ connects two or more nodes with arbitrary types. We define following three types of hyperedges to represent multi-type relations among the various nodes: (1) a \textit{feature hyperedge} connects all features of an item; (2) a \textit{price hyperedge} connects the price nodes of all the items in a session. (3) a \textit{session hyperedge} connects the ID nodes of all the items in a session. We utilize feature hyperedges to propagate embeddings of heterogeneous nodes. The price and session hyperedges are used to extract the user's price and interest preferences, respectively. Two nodes are adjacent if they are connected by a hyperedge. The process of constructing the heterogeneous hypergraph is illustrated in the leftmost part of Figure~\ref{CoHHN}.

\section{The proposed method}

\subsection{Overview of the proposed \baby}
The proposed \baby is illustrated in Figure~\ref{CoHHN}. Based ob the constructed heterogeneous hypergraph, we devise a dual-channel aggregating mechanism to propagate various information among nodes in two channels, \ie intra-type channel and inter-type channel. After that, we utilize attention layers to extract the user's price and interest preferences based on price and session hyperedges, respectively. A co-guided learning schema is then proposed to model the complex relations between these two preferences in determining the user choices. Finally, we make prediction of user actions based on item features and her price and interest preferences. Next, we elaborate the proposed model \baby.

\subsection{Dual-channel Aggregating Mechanism}
The customized heterogeneous hypergraph endows the proposed \baby with the potential to model the complex high-order dependencies among heterogeneous nodes. However, it is hard to extract useful information from heterogeneous nodes. In the constructed heterogeneous hypergraph, a target node can be adjacent with nodes from many types. It is obvious that nodes with same type contain homogeneous information, and different types have different semantics. Therefore, we summarize the message aggregation into two channels, \ie \textit{intra-type channel} and \textit{inter-type channel}. In the intra-type channel, for a target node, adjacent nodes with the same type have different importance. Also, in the inter-type channel, different types may have different impacts on the target node. To aggregate rich information from two channels, we devise a novel dual-channel aggregating mechanism.

\paratitle{Intra-type aggregating.} $\mathbf{v}^{t} \in \mathbb{R}^d$ is the embedding of a target node with type $t$. Its adjacent nodes with a certain type ($\tau$) form a set $ N_{t}^{\tau}$. The intra-type aggregating aims to distinguish the importance of nodes in $ N_{t}^{\tau}$ and aggregate the information from these nodes to represent the type ($\tau$). Formally, we learn an representation of type $\tau$ for a target node $\mathbf{v}^{t}$ as follows,
\begin{align}
    \mathbf{e}_{t}^{\tau} &= \sum_{i} \alpha_i \mathbf{v}_i^{\tau}\\
    \alpha_i &= \frac{exp(\mathbf{u}_{\tau} \mathbf{v}_i^{\tau})}{\sum_{\mathbf{v}_i^{\tau} \in N_{t}^{\tau}}\exp(\mathbf{u}_{\tau} \mathbf{v}_i^{\tau})}
\end{align}
where $\mathbf{u}_{\tau}^{\top} \in \mathbb{R}^d$ is the attention vector which determines the relevance of a node $\mathbf{v}_i^{\tau} \in N_{t}^{\tau}$ on $\mathbf{v}^{t}$, and $\mathbf{e}_{t}^{\tau} \in \mathbb{R}^{d}$ is the type embedding for $\mathbf{v}^{t}$ about type $\tau$. Considering that the nodes with same type contain homogeneous information, the intra-type aggregating merges the information at the embedding level. It is worth noting that we apply different attention vectors for different types to extract the type information, which enables our model to learn customized type embeddings for every node. We can abstract the process of intra-type aggregating defined by Eq.(3-4) as a function,
\begin{align}
    \mathbf{e}_{t}^{\tau} &= f_{a}(N_{t}^{\tau})
\end{align}

\paratitle{Inter-type aggregating.}
Based on the learned type embedding $\mathbf{e}_{t}^{\tau}$, the inter-type aggregating propagates the embeddings of different types to the target node. Given that different types provide heterogeneous information for the target node, the inter-type aggregating merges the information at the feature level as follows,
\begin{align}
    \mathbf{g}_{t} &= \sigma (\mathbf{W}_{t}[\mathbf{v}^t;\mathbf{e}_{t}^{\tau 1};\mathbf{e}_{t}^{\tau 2}] + \mathbf{W}_{t}^{\tau 1} \mathbf{e}_{t}^{\tau 1} + \mathbf{W}_{t}^{\tau 2} \mathbf{e}_{t}^{\tau 2}) \\
    \mathbf{h}^t &= \mathbf{v}^t + \mathbf{g}_{t} * \mathbf{e}_{t}^{\tau 1} + (1-\mathbf{g}_{t}) * \mathbf{e}_{t}^{\tau 2}
\end{align}
where $\mathbf{W}_{t}\in \mathbb{R}^{d \times 3d}, \mathbf{W}_{t}^{\tau 1} $ and $\mathbf{W}_{t}^{\tau 2} \in \mathbb{R}^{d \times d} $ are learnable parameters, [;] is concatenation, $ \sigma $ is sigmoid function and $*$ is element-wise product. $\mathbf{h}^t \in \mathbb{R}^{d}$ is the target node's embedding whose semantics are enriched by its adjacent nodes. The process of inter-type aggregating as illustrated in Eq.(6-7) can be defined as follows,
\begin{align}
    \mathbf{h}^t &= f_{b}(\mathbf{v}^t, \mathbf{e}_{t}^{\tau 1}, \mathbf{e}_{t}^{\tau 2})
\end{align}

With the dual-channel aggregating mechanism, we update the embeddings of all nodes as follows,
\begin{align}
    \mathbf{h}^{id} &= f_{b}(\mathbf{v}^{id}, f_{a}(N_{id}^p), f_{a}(N_{id}^c)) + avg(N_{id}^{id}) \\
    \mathbf{h}^{p} &= f_{b}(\mathbf{v}^{p}, f_{a}(N_{p}^{id}), f_{a}(N_{p}^c)) \\
    \mathbf{h}^{c} &= f_{b}(\mathbf{v}^{c}, f_{a}(N_{c}^p), f_{a}(N_{c}^{id})) 
\end{align}
where $\mathbf{v}^{id}, \mathbf{v}^p$ and $\mathbf{v}^c \in \mathbb{R}^{d} $ and $N_{id}^{id}$ consists of ID nodes adjacent with $\mathbf{v}^{id}$. The $avg(N_{id}^{id})$ calculates the averaged ID embeddings in $N_{id}^{id}$. We iteratively repeat the dual-channel aggregating operation $r$ times to fully extract the useful information from various nodes. As a result, our proposed \baby is able to model complex high-order dependencies among various features of items and capture price and interest preferences of users. Note that, we mainly use feature hyperedges to propagate information in the dual-channel aggregating mechanism. At last, we obtain embeddings for all nodes as $\mathbf{h}^{id}, \mathbf{h}^p$ and $\mathbf{h}^c \in \mathbb{R}^{d} $.

\subsection{Preferences Extracting}
Based on the learned node embeddings, we further extract users' price and interest preferences for recommendation.

\paratitle{Extracting price preferences.}
We rely on the price hyperedge, \ie $ [\mathbf{h}_1^{p},  \mathbf{h}_2^{p},  \mathbf{h}_3^{p}, ...,  \mathbf{h}_m^{p}]$, to extract the user's price preferences, given that the user's price preferences are reflected in the price of items she has purchased. Self-attention has been demonstrated to be good at capturing item-item transition patterns across the entail input sequence\cite{Attention,GC-SAN}. 
Therefore, we use self-attention with multi-head mechanism to model the price preferences as follows,
\begin{align}
    \mathbf{E}_{p} &= [\mathbf{h}_1^{p};\mathbf{h}_2^{p}; ...; \mathbf{h}_m^{p}] \\
    head_i &= Attention(W_i^Q\mathbf{E}_p, W_i^K\mathbf{E}_p, W_i^V\mathbf{E}_p) \\
    \mathbf{S}_p &= [head_1; head_2; ...; head_h]
\end{align}
where $W_i^Q, W_i^K$ and $W_i^V \in \mathbb{R}^{\frac{d}{h} \times d} $ are parameter matrices used to map the input to query, key and value respectively, and $h$ is the number of heads in the self-attention block. The hidden state $\mathbf{S}_p$ represents original price preferences of a user $\mathbf{\hat{P}}_p \in \mathbb{R}^d$,
\begin{align}
     \mathbf{\hat{P}}_p &= \mathbf{S}_p^{(t)}
\end{align}

\paratitle{Extracting interest preferences.}
We then extract the user's interest preferences based on the session hyperedge $ [ \mathbf{h}_1^{id},  \mathbf{h}_2^{id}, \mathbf{h}_3^{id}, ...  \mathbf{h}_m^{id}]$. The user's interest preferences drift along with time \cite{Zhou@SIGIR2021}. Therefore, we integrate the position embeddings with the learned ID embeddings to model the dynamic interest of users. Specifically, we utilize the reversed position embeddings \cite{GCE-GNN,DHCN} to encode the position information, \ie $\mathbf{pos}_i \in \mathbb{R}^{d}$. And then, we merge the $\mathbf{h}_i^{id}$ and $\mathbf{pos}_i$ as follows,

\begin{align}
     \mathbf{v}_i^*  = tanh(\mathbf{W}_f[\mathbf{h}_i^{id};\mathbf{pos}_i] + \mathbf{b}_f)
\end{align}
where $\mathbf{W}_f \in \mathbb{R}^{d \times 2d}$ and $\mathbf{b}_f$ are learnable parameters. $\mathbf{v}_i^* \in \mathbb{R}^d$ is the embedding of $i$-th item in the session. Following \cite{STAMP,SR-GNN}, we obtain the original interest preferences as follows,
\begin{align}
    \mathbf{\hat{I}}_p &= \sum^{t}_{i=1} \beta_i \mathbf{h}_i^{id} \\
    \beta_i &= \mathbf{u}\sigma(\mathbf{A}_1\mathbf{v}_i^{*}+\mathbf{A}_2\mathbf{\bar{v}}^{*}+\mathbf{b})
\end{align}
where $\mathbf{A}_1$, $ \mathbf{A}_2  \in \mathbb{R}^{d \times d}$ and $\mathbf{b}$ are learnable parameters, $\mathbf{u}^T \in \mathbb{R}^d$ is the attention vector and $\mathbf{\bar{v}}^{*} = \frac{1}{m}\sum_{i=1}^m \mathbf{v}_i^*$. $\mathbf{\hat{I}}_p \in \mathbb{R}^d$ is the representation for original interest preferences of a user.

\subsection{Co-guided Learning Schema}
In a session, we have learned the user's original price preferences $\mathbf{\hat{P}}_p$ and interest preferences $\mathbf{\hat{I}}_p$. Given an item $x_i$ with its ID embedding $\mathbf{v}_i^{id}$ and price embedding $\mathbf{v}_i^p$, we can simply define the probability the user will interact with it as follows,

\begin{align}
    y_i=\mathbf{\hat{P}}_p^{\top}  \mathbf{v}_i^p + \mathbf{\hat{I}}_p^{\top} \mathbf{v}_i^{id}
\end{align}

However, as stated in former sections, the price preferences and interest preferences influence each other and jointly determine users' choices. Obviously, simple addition is unable to tackle such a complex situation. Thus, we propose a novel co-guided learning schema to model the mutual relations between price and interest preferences in determining user choices.

We first integrate the price preferences $\mathbf{\hat{P}}_p$ with interest preferences $\mathbf{\hat{I}}_p$ in two ways,
\begin{align}
    \mathbf{m}_c &= tanh(\mathbf{W}_1^{pi}(\mathbf{\hat{P}}_p*\mathbf{\hat{I}}_p) + \mathbf{b}_c)\\
    \mathbf{m}_j &= tanh(\mathbf{W}_2^{pi}(\mathbf{\hat{P}}_p+\mathbf{\hat{I}}_p) + \mathbf{b}_j)
\end{align}
where $\mathbf{W}_1^{pi}$, $\mathbf{W}_2^{pi} \in \mathbb{R}^{d \times d}$, $b_c$ and $b_j$ are learnable parameters. The $\mathbf{m}_c$ and $\mathbf{m}_j \in \mathbb{R}^d$ represent interactive relations between price and interest preferences in different semantic spaces. And then, we utilize gating mechanism to further model the mutual relations between $\mathbf{\hat{P}}_p$ and $\mathbf{\hat{I}}_p$ as follows,
\begin{align}
    \mathbf{r}_P &= \sigma(\mathbf{W}_1^{r}\mathbf{m}_c + \mathbf{U}_1^{r}\mathbf{m}_j)\\
    \mathbf{r}_I &= \sigma(\mathbf{W}_2^{r}\mathbf{m}_c + \mathbf{U}_2^{r}\mathbf{m}_j)\\
    \mathbf{m}_P &= tanh(\mathbf{W}^{P}(\mathbf{r}_P*\mathbf{\hat{P}}_p) + \mathbf{U}^P((1-\mathbf{r}_I)*\mathbf{\hat{I}}_p))\\
    \mathbf{m}_I &= tanh(\mathbf{W}^{I}(\mathbf{r}_I*\mathbf{\hat{I}}_p) + \mathbf{U}^I((1-\mathbf{r}_P)*\mathbf{\hat{P}}_p))
\end{align}
where $\mathbf{W}_1^{r}$,$\mathbf{W}_2^{r}$, $\mathbf{W}^{P}$, $\mathbf{W}^{I}$, $\mathbf{U}_1^{r}$, $\mathbf{U}_2^{r}$, $\mathbf{U}^{P}$ and $\mathbf{U}^{I} \in \mathbb{R}^{d \times d}$. The $\mathbf{r}_P$ ( $\mathbf{r}_I$ ) $ \in \mathbb{R}^d$ represents 'the remember gate' which controls how much price (interest) preferences are retained when modeling the relations between them. Note that we also apply the complementary of $\mathbf{r}_I$ ($\mathbf{r}_P$), \ie $ 1 - \mathbf{r}_I$ ($ 1 - \mathbf{r}_P$), to incorporate interest (price) preferences to guide the learning of price (interest) preferences, which enriches the semantics of these two preferences. Finally, we obtain the user's price and interest preferences as follows,
\begin{align}
    \mathbf{P}_p &=\mathbf{m}_P * ( \mathbf{\hat{P}}_p + \mathbf{m}_I)\\
    \mathbf{I}_p &= \mathbf{m}_I * ( \mathbf{\hat{I}}_p + \mathbf{m}_P)
\end{align}
where $\mathbf{P}_p$ and $\mathbf{I}_p \in \mathbb{R}^d$ are the user's price preferences and interest preferences whose semantics are enriched by each other. Note that the price and interest preferences ($\mathbf{\hat{P}}_p$ and $\mathbf{\hat{I}}_p$) extract information from each other to guide the learning process, which enables our \baby to model the complex relations between them in determining user choices.

\subsection{Prediction and Training}
Given a session, we can calculate the likelihood score $y_i$ corresponding to all items $x_i$ $\in$ $\mathcal{I}$ based on the user's preferences ($\mathbf{P}_p$ and $\mathbf{I}_p$) and item's features ($\mathbf{v}_i^{id}$ and $\mathbf{v}_i^p$) as follows, 
\begin{align}
    y_i &= \mathbf{P}_p^{\top}  \mathbf{v}_i^p + \mathbf{I}_p^{\top}  \mathbf{v}_i^{id}
\end{align}

Then we process it with softmax function to obtain the final score as follows,
\begin{align}
    y_i &=  \frac{exp(y_i)}{\sum_{j=1}^nexp(y_j)}
\end{align}

Our model can be trained by using cross-entropy loss as follows:
\begin{align}
    \mathcal{L}(\mathbf{p}, \mathbf{y}) = - \sum^n_{j=1} p_j \log (y_j) + (1-p_j)\log(1-y_j)
\end{align}
where $p_j$ is the ground truth that indicates whether the user clicks on item $x_j$ and $y_j$ is the predicted probability of item $x_i$ to be interacted with next. 

\section{Experimental Setup}
\subsection{Research Questions}
We conduct experiments on three real-world datasets for session-based recommendation to evaluate our proposed \baby, with the purpose of answering following research questions: 

\begin{itemize}
    \item \textbf{RQ1} Does the proposed model \baby achieve state-of-the-art performance? (see Section 6.1)
    
    \item \textbf{RQ2} What is the effect of the price factor in SBR? (see Section 6.2)
    
    \item \textbf{RQ3} What is the effect of the co-guided learning schema? (see Section 6.3)
    
    \item \textbf{RQ4} How well does the \baby perform at different price levels?  (see Section 6.4)
    
    \item \textbf{RQ5} What is the influence of the key hyper-parameters on \baby? (see Section 6.5)
\end{itemize}

\subsection{Datasets and Preprocessing}
\begin{table}[t]
\tabcolsep 0.05in 
\centering
\caption{Statistics of the datasets.}
\begin{tabular}{ccccc}
\toprule
Datasets      &Cosmetics & Diginetica-buy & Amazon \\
\midrule
\#item        & 23,194   & 24,889    & 9,114     \\
\#price level       & 10       & 100       & 50        \\
\#category    & 301      & 721       & 613      \\
\#interaction & 1,058,263  & 855,070 & 487,701    \\
\#session     & 156,922  & 187,540   & 204,036     \\
avg.length    & 6.74     & 4.56      & 2.39      \\
\bottomrule
\end{tabular}

\label{statistics}
\end{table}

We evaluate the proposed \baby and all baselines on following three real-world public datasets:
\begin{itemize}
    \item Cosmetics \footnote{\url{https://www.kaggle.com/mkechinov/ecommerce-events-history-in-cosmetics-shop}} is a kaggle competition dataset, which records user behaviors in a medium cosmetics online store. We use one month (October 2019) records and only retain the interactions with type 'add\_to\_cart' or 'purchase' in our work. 
    
    \item Diginetica-buy \footnote{\url{https://competitions.codalab.org/competitions/11161}} is an e-commerce dataset which is composed of user purchasing behaviors on the websites. 
    
    \item Amazon \footnote{\url{http://jmcauley.ucsd.edu/data/amazon/}}, as a widely used  benchmark dataset for recommendation, contains users' purchase behavior from Amazon \cite{Amazon}. One representative sub-dataset 'Grocery and Gourmet Food' is adopted in our experiments. We view sequential actions within one day of each user as a session. 
    
\end{itemize}

Following \cite{Zheng@ICDE2020, NARM,SR-GNN,LESSR,DHCN,Xia@CIKM2021}, the sessions of length 1 and items that appear less than 10 times are filtered out in all datasets. For a session, the last item is viewed as the label and the remaining sequence is used to model user preferences. We use the earliest 70\% (chronologically) of the sessions as the training set and use the next 20\% of sessions as the validation set for hyper-parameter tuning on \baby and all baselines. The remaining 10\% constitutes the test set for reporting model performance. Statistical details of three datasets are shown in Table~\ref{statistics}.

\subsection{Evaluation Metrics} 
As in \cite{GRU4Rec,NARM,BERT4Rec,SR-GNN,LESSR,DHCN,Xia@CIKM2021}, we use following two common metrics to evaluate the performance of all methods. 
\begin{itemize}
\item Prec@k: Precision measures the proportion of cases in which the ground-truth item is within the recommendation list. 
\item MRR@k: Mean Reciprocal Rank is the average of reciprocal ranks of the ground-truth items among the recommendation lists. 
\end{itemize}
Note that, the Prec@k does not consider the rank of the item as long as it is amongst the recommendation list. While the MRR@k takes the rank of the item into account, which is important in setting where the order of recommendations matters. 
And larger metric values indicate better performances for both Prec@k and MRR@k. 
In this work, we report the results with $k$ = 10, 20.

\subsection{Baselines}
To evaluate the performance of the proposed model \baby, we choose following competitive methods as baselines:
\begin{itemize}
    \item \textbf{S-POP} recommends the most frequent items in the current session.  
    
    \item \textbf{SKNN} computes the scores of candidate items according to their occurrences in the neighbor sessions. 
    
    \item \textbf{GRU4Rec} \cite{GRU4Rec} utilizes GRU to mine the sequential patterns within sessions. 
    \item \textbf{NARM} \cite{NARM}  employs RNNs with attention mechanism to capture the user’s main purpose. 
    
    \item \textbf{BERT4Rec} \cite{BERT4Rec} employs a bidirectional self-attention architecture to encode historical interaction sequences. 
    
    \item \textbf{SR-GNN} \cite{SR-GNN} builds session graph and utilizes graph neural networks to capture the pairwise transition between items.
    
    \item \textbf{LESSR} \cite{LESSR} tackles information loss of GNN-based models in SBR by introducing shortcut graph attention and edge-order preserving aggregation layers.
    
    \item \textbf{$S^2$-DHCN}  \cite{DHCN} captures beyond-pairwise relations among items by hypergraph and uses self-supervised learning to enhance performance.
    
    \item \textbf{COTREC} \cite{Xia@CIKM2021} combines self-supervised learning with co-training to mine informative self-supervision signals for SBR.
    
\end{itemize}

\subsection{Implementation Details}
For fair comparison, the embedding size is set to 128 for all models. The hyper-parameters of \baby and all baselines are optimized via the grid search on the validation set. For \baby, the number of self-attention heads is 4 ($h=4$); the number of price levels (\ie $\rho$) is 10, 100 and 50 for Cosmetics, Diginetica-buy and Amazon respectively; the iterative number of dual-channel aggregating (\ie $r$) is 3 for Cosmetics/Diginetica-buy and 2 for Amazon. The model optimization is performed by using Adam with the initial learning rate $0.001$, and the mini-batch size is fixed at $100$. As in the methods~\cite{NARM}, we also truncate Back-Propagation Through Time (BPTT) at $19$ time steps. We have released the source code and datasets of our model online.\footnote{{https://github.com/Zhang-xiaokun/CoHHN}}

\begin{table*}[ht]
\small
\tabcolsep 0.04in 
  \centering
    \caption{Performance comparison of \baby with baselines over three datasets. The results (\%) produced by the best baseline and the best performer in each column are underlined and boldfaced respectively. Statistical significance of pairwise differences of \baby against the best baseline (*) is determined by the t-test ($p < 0.01$).}
    \begin{tabular}{c cccc cccc cccc}  
    \toprule  
    \multirow{2}*{Method}& 
    \multicolumn{4}{c}{Cosmetics}&\multicolumn{4}{c}{Diginetica-buy}&\multicolumn{4}{c}{Amazon}\cr  
    \cmidrule(lr){2-5} \cmidrule(lr){6-9} \cmidrule(lr){10-13}
    &Prec@10&MRR@10&Prec@20&MRR@20 &Prec@10&MRR@10&Prec@20&MRR@20 &Prec@10&MRR@10&Prec@20&MRR@20\cr  
    \midrule  
    S-POP           &32.83&26.63&38.43&27.32    &25.51&18.82&25.91&19.84    &34.60&31.96&38.03&32.19\cr  
    SKNN            &40.22&30.40&47.63&30.80    &45.68&20.24&55.76&21.10    &61.55&46.07&64.23&46.30\cr  
    GRU4Rec         &19.41&14.43&21.80&14.60    &22.04&11.32&27.88&11.73    &55.43&51.43&56.41&51.70\cr  
    NARM            &42.63&34.17&46.29&34.52    &46.56&21.76&57.34&23.27    &63.21&57.07&65.38&57.23\cr
    Bert4Rec        &38.52&23.38&46.30&23.92    &47.50&20.44&60.89&21.51    &62.83&56.57&64.52&56.74\cr  
	SR-GNN          &44.11&34.59&48.01&34.96    &45.74&21.32&56.80&22.87    &63.52&\underline{57.46}&65.83&\underline{57.89}\cr
	LESSR           &38.80&24.45&46.32&24.97    &47.88&20.82&61.35&22.64    &62.48&56.53&64.18&56.69\cr
	$S^2$-DHCN      &40.48&32.86&47.95&33.13    &45.89&21.08&54.91&22.03    &58.67&49.86&60.47&50.03\cr
	COTREC          &\underline{45.11}&\underline{34.83}&\underline{50.12}&\underline{35.24}    &\underline{48.14}&\underline{22.07}&\underline{61.82}&\underline{23.75}    &\underline{64.01}&50.06&\underline{66.44}&50.25\cr
    \midrule  
	{\bf CoHHN}     &{ $ \bf 47.88^*$ }&{$\bf 36.38^*$}&{$\bf 53.56^*$ }&{$\bf 36.79^*$ }
	                &{ $ \bf 50.57^*$ }&{$\bf 24.81^*$}&{$\bf 64.02^*$ }&{$\bf 25.76^*$ }
	                &{ $ \bf 65.32^*$ }&{$\bf 58.78^*$}&{$\bf 67.69^*$ }&{$\bf 59.01^*$ }\cr
	\bottomrule
    \end{tabular}
    \label{performance}
\end{table*}

\section{Results and Analysis}

\subsection{Overall Performance (RQ1)}

Table~\ref{performance} shows the comparison results of \baby over other competitive approaches, where the following observations are noted: 

(1) Among baselines, different methods perform differently on different datasets. All of the baselines aim at modeling single type of information to capture users' interest preferences, while ignoring that other factors, \eg price, can also affect user choices. This leads to that the baselines are incapable of accurately predicting user actions across different datasets.

(2) NARM and BERT4Rec outperform GRU4Rec by a large margin, which can own to the attention mechanism they introduce for capturing the main intents in a session. However, BERT4Rec does not perform as good as expected, considering that it has achieved tremendous success in the task of natural language processing. The main advantage of BERT4Rec's structure is that it can capture distant dependencies among items within sequences. As shown in the Table~\ref{statistics}, the length of sequence is generally short in session data. We argue that this is the reason for its poor performance.

(3) Relying on the ability for modeling pairwise relations between nodes, GNN-based methods (SR-GNN and LESSR) obtain competitive performance. COTREC further improves the performance of GNN by constructing session graphs in two views. We argue that the view augmentation of COTREC enables the model to capture beyond-pairwise relations among items. Also, $S^2$-DHCN utilizes the unique structure of hypergraph to do so. However, in essence, the above methods can only capture the single type of information, while hard to model the price and interest preferences hidden in the various features of items. Therefore, its performance is still inferior to the proposed \baby.

(4) Our proposed \baby shows consistent superiority over all baselines in terms of all evaluation metrics on all datasets. Specifically, \baby outperforms best performer in Prec@20 and MRR@20 by 6.86\% and 4.40\% on Cosmetics, 3.56\% and 8.46\% on Diginetica and 1.88\% and 1.93\% on Amazon. We argue that the improvements of the proposed \baby benefit from taking price preferences into account. To model the user's price and interest preferences, we devise a novel heterogeneous hypergraph network which is able to capture complex high-order dependencies among various features of items. Besides, the co-guided learning schema models the relations between price preferences and interest preferences in users' decision-making process, helping the model accurately formulate user future actions.

\subsection{The Effect of the Price Factor (RQ2)}

\begin{table}[t]
\small
\tabcolsep 0.02in 
  \centering
    \caption{Performance of variants of \baby .}
    \begin{tabular}{c cc cc cc}  
    \toprule  
    \multirow{2}*{variants}& 
    \multicolumn{2}{c}{Cosmetics}&\multicolumn{2}{c}{Diginetica-buy}&\multicolumn{2}{c}{Amazon}\cr  
    \cmidrule(lr){2-3} \cmidrule(lr){4-5} \cmidrule(lr){6-7}
    &Prec@20&MRR@20 &Prec@20&MRR@20 &Prec@20&MRR@20\cr  
    \midrule  
    COTREC      &50.12&35.24    &61.82&23.75    &66.44&50.25\cr
    CoHHN-c     &53.05&36.28    &63.21&24.97    &67.06&58.51\cr  
    CoHHN-p     &52.81&35.92    &61.83&22.99    &66.93&58.12\cr
    CoHHN-pp    &52.59&36.08    &62.30&23.23    &67.14&58.38\cr
	{\bf CoHHN} &{ $ \bf 53.56^*$ }&{$\bf 36.79^*$} 
	            &{$\bf 64.02^*$ }&{ $ \bf 25.76^*$ }
	            &{$\bf 67.69^*$}&{$\bf 59.01^*$ }\cr
	\bottomrule
    \end{tabular}

    \label{priceEffect}
\end{table}  

\begin{figure*}[ht]
  \centering
  \includegraphics[width=0.85\linewidth]{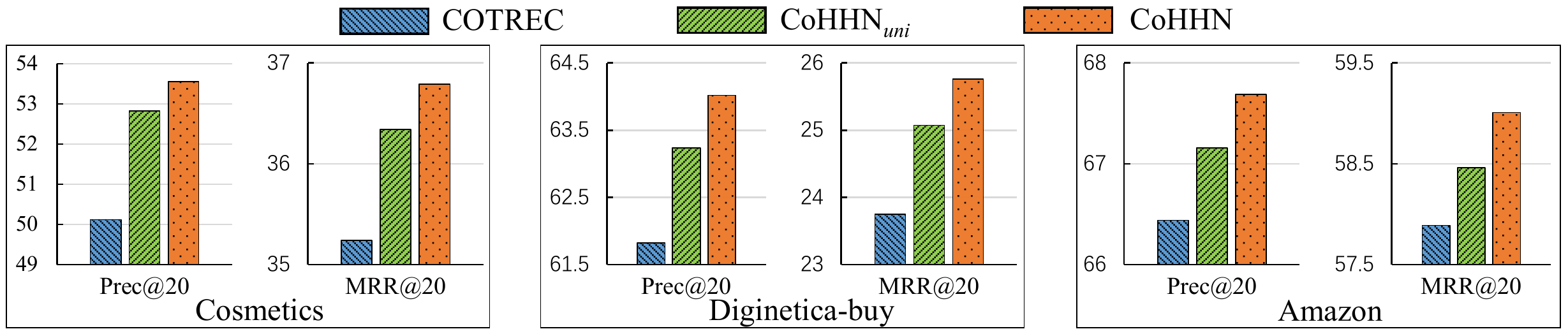}
  \caption{The Effect of the price discretization}\label{logistics}
\end{figure*}

\begin{figure*}[ht]
  \centering
  \includegraphics[width=0.85\linewidth]{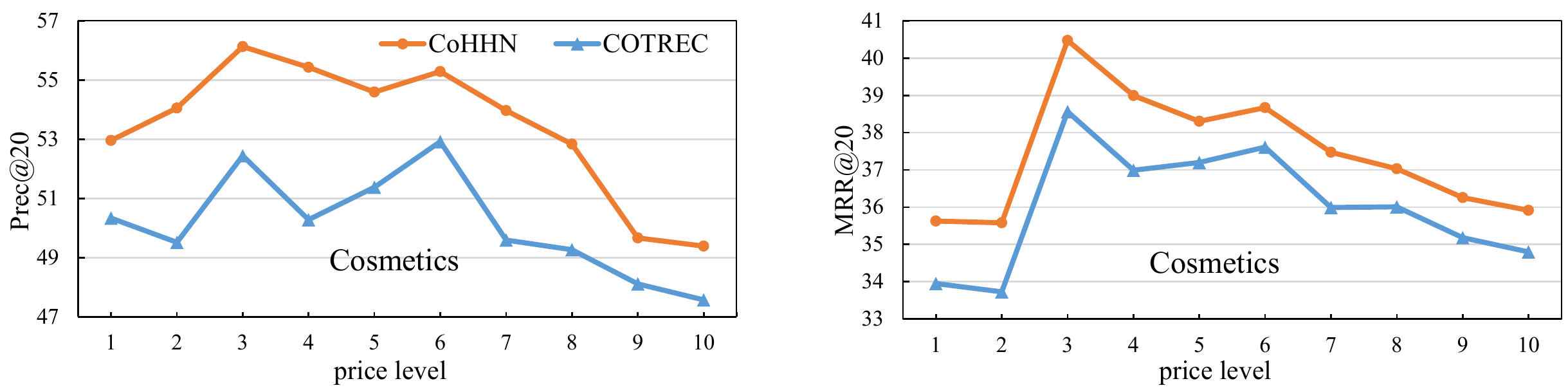}
  \caption{Performance on different price levels in Cosmetics }\label{priceLevel}
\end{figure*}

A key novelty of \baby is that it incorporates the price factor and models both price and interest preferences to predict user actions in SBR. Moreover, we also incorporate item categories into \baby, considering that users' price preferences vary widely across different categories. \baby-p and \baby-c represent the versions of \baby without considering price and category respectively. And \baby-pp only incorporates the price information to update node embeddings but does not explicitly extract users' price preferences. As can be observed in Table~\ref{priceEffect}, \baby-pp outperforms \baby-p and \baby-c performs better than \baby-p, which indicates that incorporating price into SBR does improve the recommendation accuracy. It is also in line with intuition that price plays a critical role in user purchase process. \baby achieves large improvements over \baby-pp. It demonstrates that price preferences should be considered as a decisive factor rather than auxiliary one when predicting user future actions. \baby also defeats \baby-c, which reveals that items' category contributes to predicting users' choices. We argue that the introduction of category information enables the \baby to learn users' price preferences more precisely.

In order to encode price information across different categories, we discretize absolute price into different price levels by equally dividing logistic probability distribution. We demonstrate the effectiveness of this design choice with the variant: \babyuni$_{uni}$ discretizes price into different price levels via uniform quantization as in \cite{Zheng@ICDE2020}, \ie $ p_i = \lfloor \frac{x_p - min}{max - min} \times \rho \rfloor $. As shown in Figure~\ref{logistics}, \baby performs better than the \babyuni$_{uni}$. It suggests that our method could encode price information more accurate and reveals the price preferences of users. Furthermore, \babyuni$_{uni}$ achieves a large performance margin over the best baseline COTREC, which indicates the significance of incorporating price preferences in the task again.

\subsection{The Effect of the Co-guided Learning (RQ3)}
\begin{table}[t]
\small
\tabcolsep 0.02in 
  \centering
    \caption{Effect of the Co-guided Learning}
    \begin{tabular}{c cc cc cc}  
    \toprule  
    \multirow{2}*{variants}& 
    \multicolumn{2}{c}{Cosmetics}&\multicolumn{2}{c}{Diginetica-buy}&\multicolumn{2}{c}{Amazon}\cr  
    \cmidrule(lr){2-3} \cmidrule(lr){4-5} \cmidrule(lr){6-7}
    &Prec@20&MRR@20 &Prec@20&MRR@20 &Prec@20&MRR@20\cr  
    \midrule  
    COTREC      &50.12&35.24    &61.82&23.75    &66.44&50.25\cr
    CoHHN-co     &52.79&36.33   &62.49&23.54   &67.28&58.63\cr  
	{\bf CoHHN} &{ $ \bf 53.56^*$ }&{$\bf 36.79^*$} 
	            &{$\bf 64.02^*$ }&{ $ \bf 25.76^*$ }
	            &{$\bf 67.69^*$}&{$\bf 59.01^*$ }\cr
	\bottomrule
    \end{tabular}

    \label{Co-guided}
\end{table}

The price and interest preferences influence each other and jointly determine user behaviors. Therefore, we propose a co-guided learning schema to model the complex relations between these two preferences in determining user choices. \baby-co uses simple addition to calculate the probability of interactions, that is, as formulated by Eq.(19). According to Table~\ref{Co-guided}, \baby defeats \baby-co in a large margin. It indicates that the relations between price and interest preferences in determining user choices are complex and the simple addition can not handle the complex situation. The proposed co-guided learning schema is able to model the mutual relations between price and interest preferences, thus improving the performance for SBR.

\subsection{Performance in Different Price Levels(RQ4)}

\begin{figure*}[ht]
  \centering
  \includegraphics[width=0.9\linewidth]{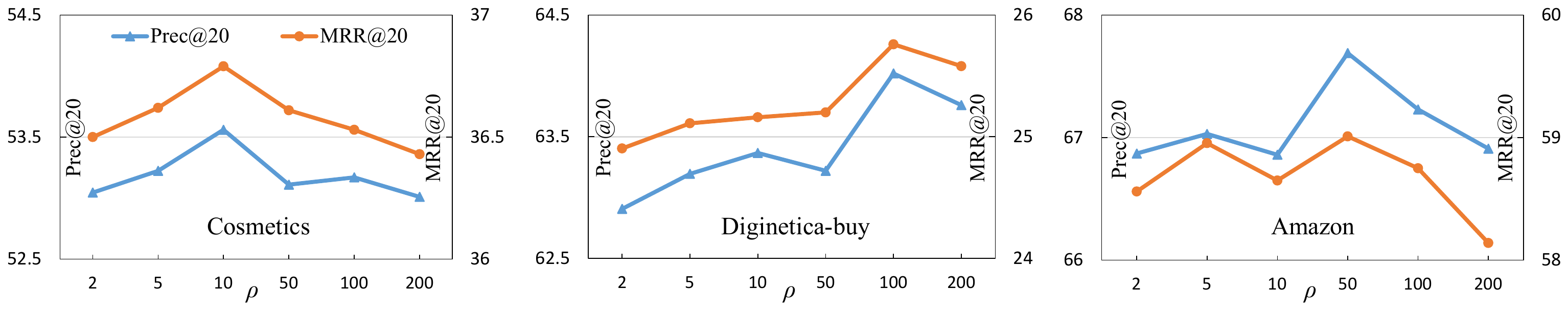}
  \caption{Performance of \baby on different number of price levels}\label{priceNumber}
\end{figure*}

\begin{figure*}[ht]
  \centering
  \includegraphics[width=0.9\linewidth]{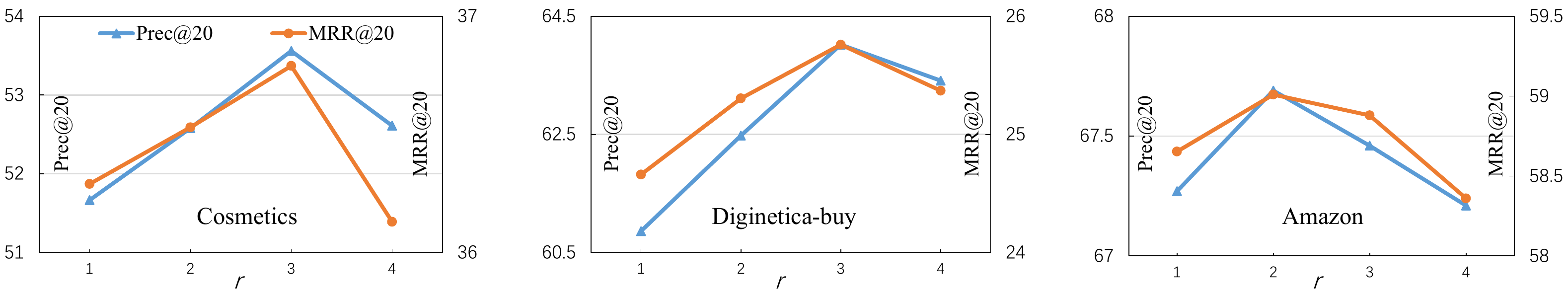}
  \caption{Performance of \baby on different iterative number of the dual-channel aggregating}\label{iterativeNumber}
\end{figure*}

As demonstrated in the previous sections, price factor has a great impact on users' behaviors. Modeling price and interest preferences can improve the accuracy of predicting user actions. Nevertheless, users show different behavior characteristics in these two preferences. For interest preferences, users present huge personalized differences. As for price preferences, most users have a bias towards low prices. Therefore, average accuracy might increase if we blindly recommend low-price items to users. However, such a method may decrease the revenue of manufacturers, because high-price items often bring high profits. It requires us to improve accuracy in not only low-price items but also high-price items, when incorporating price factor into SBR. Therefore, we report the results of \baby and the best baseline COTREC under different price levels. Due to space limit, only the results on Cosmetics are presented, while we obtained similar observations on other datasets.

As shown in Figure~\ref{priceLevel}, \baby outperforms COTREC at all price levels on both metrics. It demonstrates that our \baby is able to percept users' price preferences at different price levels and make personalized recommendation accordingly. The consistent improvements over all price levels ensure that our \baby can help manufacturers' earning to grow while improving the prediction accuracy. The method we use to discretize price can optimize price allocation. We speculate that this is one of the reasons why our model works well in all price levels. Furthermore, \baby and COTREC have high accuracy for items with mid and low price levels (\ie 3-6). It is in line with the facts that most people tend to purchase items with mid and low prices.

\subsection{Hyper-parameter Study (RQ5)}
In this section, we evaluate the influence of two key hyper-parameters on the proposed \baby, \ie the number of price levels $\rho$ and the iterative number of the dual-channel aggregating $r$.

The number of price levels $\rho$ determines the granularity of the price preferences modeling. We conduct experiments on three datasets under different values of $\rho$ to study its influence on model's performance. The performance patterns are plotted in the Figure~\ref{priceNumber}. When the $\rho$ is set extremely low like $\rho = 2$ which means the model only roughly classifies price as expensive or cheap, the model is unable to learn the various price preferences, thus the performance is unsatisfactory. While if we set the $\rho$ too high as 200 in the Cosmetics, items with near price are assigned to different price levels. In this case, there is no significant difference between different price levels, causing performance degradation. In absolute terms, the values close to the optimal $\rho$ present similar performance. We argue that the stable performance benefits from our proposed method for price discretization. The method of equal probability segmentation makes the price reasonably arranged into each level, thus contributing to learning the user's price preferences. Furthermore, the optimal value for $\rho$ is different in different datasets, \ie 10 in Cosmetics, 100 in Diginetica-buy and 50 in Amazon. We speculate that the diversity of users' price preferences on different datasets results in different optimal values for $\rho$.

The iterative number of the dual-channel aggregating $r$ controls the degree of information fusion among nodes. For a node in the heterogeneous hypergraph, with each iteration, it incorporates some information of adjacent nodes. As shown in the Figure~\ref{iterativeNumber}, with the increase of $r$, the model's performance improves first, and then drops. The reason for this is that the degree of information fusion among nodes increases when $r$ becomes larger, but performing iteration too many times does not help because that the over-smooth problem occurs. Moreover, the optimal value of $r$ is different for different datasets, \ie 3 for Cosmetics/Diginetica-buy and 2 for Amazon. As shown in Table~\ref{statistics}, the Amazon contains less items than other two datasets. We believe that fewer nodes in the heterogeneous hypergraph make the model learn accurate node embeddings through fewer iterations.

\section{Conclusion and Future Work}
Price factor plays an important role in determining user purchase behaviors, which motivates us to propose a novel model Co-guided Heterogeneous Hypergraph Network (\baby) to incorporate price into session-based recommendation. Specifically, we innovatively devise a customized heterogeneous hypergraph network to model complex high-order dependencies among various features of items to mine users' price and interest preferences. In the heterogeneous hypergraph network, a dual-channel aggregating mechanism is devised to aggregate various information among heterogeneous nodes and multiple relations. Based on the learned node embeddings, we extract users' price and interest preferences within a session via attention layers. After that, we propose a co-guided learning schema to model complex relations between these two preferences in determining user choices. Comprehensive experiments conducted on three real-world datasets demonstrate the superiority of the proposed \baby over state-of-the-art methods. Further analysis also validates the significance of price factor in session-based recommendation. 

As to future work, we plan to extend \baby to incorporate more information such as item brand and seller's reputation to further explore the preferences of users. Moreover, although specifically devised for modeling price and interest preferences in session-based recommendation, our proposed heterogeneous hypergraph can be easily modified to address other tasks which need to mine knowledge from complex heterogeneous information. In addition, it is also an interesting direction to model the dynamics of users' price preferences.

\newpage

\section*{Acknowledgement}
This work has been supported by the Natural Science  Foundation  of  China (No.62076046, No.62006034), Natural Science Foundation of Liaoning Province (No.2021-BS-067) and the Fundamental Research Funds for the Central Universities (No.DUT21RC(3)015). We would like to thank the anonymous reviewers for their valuable comments.


\bibliographystyle{ACM-Reference-Format}
\bibliography{ref}

\end{document}